\documentclass[aps,prb,twocolumn,showpacs,floatfix,superscriptaddress]{revtex4-1}

\usepackage[dvips]{graphics}
\usepackage{epsfig}
\usepackage{times}
\usepackage{amsmath,amssymb}
\usepackage{amsthm}
\usepackage{subfigure,overpic}
\usepackage{mathrsfs}
\usepackage{multirow}

\DeclareMathOperator{\Tr}{Tr}

\def \beq {\begin{equation}}
\def \edq {\end{equation}}
\def \bes {\begin{subequations}}
\def \eds {\end{subequations}}
\def \beqn {\begin{equation*}}
\def \edqn {\end{equation*}}

\begin{document}
\title{Nonlinear heat transport in mesoscopic conductors: Rectification, Peltier effect
and Wiedemann-Franz law}
\author{Rosa L\'opez}
\author{David S\'anchez}
\affiliation{Institut de F\'{i}sica Interdisciplin\`{a}ria i de Sistemes Complexos
IFISC (UIB-CSIC), E-07122 Palma de Mallorca, Spain}
\affiliation{Departament de F\'{i}sica, Universitat de les Illes Balears,
E-07122 Palma de Mallorca, Spain}

% 73.23.-b -> Electronic transport in mesoscopic systems 
% 73.50.Lw -> Thermoelectric effects 
% 73.63.Kv -> Quantum dots
%  73.50.Fq % - High-field and nonlinear effects

\begin{abstract}
We investigate nonlinear heat properties in mesoscopic conductors
using a scattering theory of transport. Our approach is based
on a leading-order expansion in both the electrical and thermal
driving forces. Beyond linear response, the transport coefficients
are functions of the nonequilibrium screening potential that builds up
in the system due to interactions.
Within a mean-field approximation, we self-consistently calculate
the heat rectification properties of a quantum dot attached to two
terminals. We discuss nonlinear contributions to the Peltier effect
and find departures from the Wiedemann-Franz law in the nonlinear
regime of transport.
\end{abstract}

\maketitle

\section{Introduction} 
Much recent work has raised the interest in fundamental questions
on energy transport and heat flow at the nanoscopic scale where quantum effects
become dominant.\cite{dha08,dub11}
Mesoscopic conductors are particularly suitable for investigations
of nonequilibrium phenomena in quantum systems since electron transport
can be created, manipulated and detected thanks to the use
of electric biases or thermal gradients applied to electronic terminals
coupled to the mesoscopic sample. In this way, thermovoltages
generated in response to a temperature difference
have been observed in quantum point contacts,\cite{mol92}
quantum dots\cite{dzu97,god99} and ballistic microjunctions.\cite{mat12}
These experiments can be successfully explained using
linear-response theoretical models.\cite{but90} %,siv86,stre89,pro91}
Much more scarce
are contributions that explore the nonlinear regime of transport.%
\cite{sta93,sve12}
We have recently proposed in Ref.~\onlinecite{san12} a general scattering theory
of transport for mesoscopic {\em thermoelectric} effects
in the weakly nonlinear regime taking into account
screening interactions out of equilibrium.
A natural development of our theory would aim
at considering nonlinear effects in the {\em heat}
current flowing through a mesoscopic system.
This is the goal of the present work.

Heat currents can be driven by electric voltages
or temperature differences. In the former case,
a heat transfer accompanies an electric current, which
in the limit of low voltage gives rise to the Peltier effect;
in the latter case, a thermal gradient produces a heat flux,
which in the limit of small temperature differences
is well described by the Fourier law. Nonlinear effects in mesoscopic systems
demand the application of large driving forces across small distances,
a requirement within the scope of today's techniques.\cite{ven01}
Nonlinearities have been predicted to cause thermal rectification effects \cite{ruo09,kuo10,che08}
and low temperature cooling.\cite{whi12}
A deep understanding of nonlinear effects is also needed
for a careful assessment of the device performance 
of heat engines\cite{hum02}, heat pumps\cite{seg06,Mosk02}
and multiterminal heat-to-electric current converters.\cite{san11,ora12}
\begin{figure}
  \centering
 \includegraphics[width=0.4\textwidth, clip]{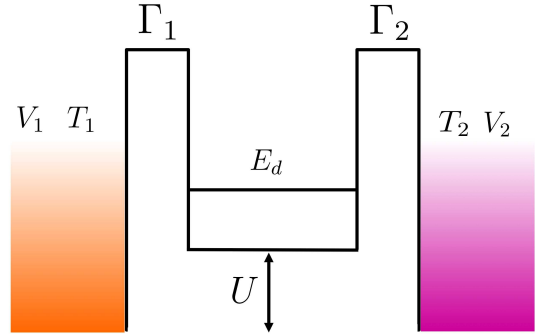}
\caption{(Color online)  Sketch of a dot system driven out of equilibrium by voltage biases 
$V_1$ and $V_2$  and thermal differences $T_1$ and $T_2$.
$\Gamma_1$ and $\Gamma_2$ denote the reservoir-dot tunneling rates
to the external reservoirs and  $E_d$ represents to the dot energy level.
$U$ is the internal dot potential illustrated as a shift
of the dot bottom band.}
  \label{fig:1}
\end{figure}
Our theory is based on a voltage and temperature expansion
around the equilibrium state. To lowest order, the linear response coefficients
depend on the electrostatic potential at equilibrium only.
Higher order terms in the expansion define weakly nonlinear transport
coefficients that are functions of the potential landscape
out of equilibrium.\cite{but93} This dependence cannot be neglected
and is crucial to formulate a gauge-invariant theory.\cite{chr96}
We determine the screening
potential up to the first order in the driving forces.\cite{san12}
In the isothermal case, charge pile-up processes are governed
by {\em particle injectivities}. Importantly, these injectivities break
the Onsager symmetry under reversal of an external magnetic field,\cite{san04}
which has been confirmed experimentally.\cite{mar06,let06,zum06,ang07,har08}
In the isoelectric case, charge can be injected to or from the system depending on whether
the carrier's energy lies above or below the chemical potential.
Thus, we define {\em entropic injectivities}\cite{san12} that specify the charge pile-up
in response to a pure thermal gradient. The electrostatic potential
is then completely determined once the bare and the screening
charges are calculated. As a result, the transmission function
becomes a function of energy, voltage and temperature shifts.\cite{san12}
Below, we discuss the consequences of this important result for
the nonlinear regime of heat transport.

We examine two applications of our theory. First, we consider
nonlinear contributions to the Peltier coefficient. Strikingly enough,
there are very few works devoted to the Peltier effect beyond
linear response. Exceptions are Ref.~\onlinecite{kul94}
on metallic constrictions, Ref.~\onlinecite{bog99}
on quantum point contacts and
Ref.~\onlinecite{zeb07} on bulk semiconductors.
We consider a quantum dot attached to two terminals
and calculate self-consistently the Peltier coefficient
for higher-order currents. Interestingly,
we find that the weakly nonlinear contributions
are expressed in terms of a ratio difference
that quantifies the relative importance of the nonlinear
conductances with respect to the linear ones. Additionally,
we test our analytical results with numerical calculations
of the full model.

Second, we examine departures of the Wiedemann-Franz
law out of equilibrium. We recall that this rule
establishes a proportionality relationship between the linear thermal
and electric conductances and that the proportionality factor,
the Lorenz number, depends on universal physical constants only.
Naturally, the Wiedemann-Franz law is not valid outside the Sommerfeld
theory of metals (noninteracting electrons and linear-response regime),
although it has been empirically found to be satisfied in bulk metals
within a wide range of temperatures.
Violations of this law in mesoscopic systems
have been investigated in artificial Kondo impurities,\cite{boe01,don02}
in the fluctuating dynamics of open quantum dots,\cite{vav05}
in single-electron transistors,\cite{kub08}
in strongly interacting dots coupled to ferromagnetic leads\cite{kra06}
and in double quantum dots.\cite{tro12}
Here, we discuss deviations from the Wiedemann-Franz law
that emerge in the nonlinear regime of transport only.
We find that these departures become maximal when the distance
between the Fermi energy and the dot level is of the order of the
level broadening because in that case the dot transmission
energy dependence becomes strongest.

\section{Theoretical model} 
Our system consists of a mesoscopic conductor attached to multiple terminals $\alpha,\beta\ldots$. Each
terminal is electrically and thermally biased as $\mu_\alpha=E_F+eV_\alpha$
and $\theta_\alpha=T_\alpha-T$,  respectively, $\mu_\alpha$ being the electrochemical potential, $E_F$ the Fermi energy,
$T_\alpha$ the temperature of lead $\alpha$ and $T$ the background temperature.
Charge and heat transport is completely characterized 
by the scattering matrix $s_{\alpha\beta}=s_{\alpha\beta}(E,eU)$. Generally, 
the scattering matrix depends on the carrier energy $E$ and the electrostatic potential inside the  sample $U$.\cite{but93,chr96}
The internal potential $U=U(\vec{r},\{V_\gamma\},\{\theta_\gamma\})$ is a function of the position $\vec{r}$ and  the set of applied voltages $\{V_\gamma\}$
and temperature gradients $\{\theta_\gamma\}$.\cite{san12}
Therefore, a complete calculation of $s_{\alpha\beta}$ as a function of the full potential landscape $U$
in the presence of interactions out of equilibrium seems an insurmountable task.
We will then focus on the weakly nonlinear transport regime for which the transport
coefficients and the system's response can be expressed in terms of quantities
evaluated at equilibrium.\cite{but93}

We denote with $\mathcal{I}_\alpha$ and $\mathcal{J}_\alpha$
the charge and heat currents, respectively, that flow from the leads
toward the sample:
\begin{eqnarray}\label{exactcurrents1}
I_\alpha =\frac{2e}{h}\sum_\beta\int dE A_{\alpha\beta}(E,eU) f_{\beta}(E)\,,
\\ \label{exactcurrents2}
\mathcal{J}_\alpha=\frac{2}{h}\sum_\beta\int dE (E-\mu_\alpha) A_{\alpha\beta}(E,eU) f_{\beta}(E)\,,
\end{eqnarray}
where $A_{\alpha\beta}=\Tr [\delta_{\alpha\beta}-
s_{\alpha\beta}^\dagger s_{\alpha\beta}]$ and  $f_\beta(E)=1/(1+\exp{[(E-E_F-eV_\beta)/k_B T_\beta]})$
is the Fermi distribution function in reservoir $\beta$. A treatment of  Eq. \eqref{exactcurrents1} in terms of interacting Green functions can be achieved following Ref. \onlinecite{Ng}. Our discussion here is entirely based on the scattering approach.

The sum over all heat flows is $\sum_{i} \mathcal{J}_\alpha=\sum_{\alpha }\mathcal{J}^E_\alpha-\sum_\alpha \mathcal{I}_\alpha V_{\alpha}$, where $\mathcal{J}^E_\alpha=(2/h)\sum_\beta\int dE E A_{\alpha\beta}(E,eU) f_{\beta}(E)$ is the energy current. 
For time-independent driving forces, the energy current is conserved\cite{mahan}
since unitarity of the scattering matrix ($\sum_{\alpha} A_{\alpha\beta}=0$) dictates that 
$\mathcal{J}^E_\alpha=0$. Heat fluxes thus satisfy the
sum rule $\sum_{\alpha} (\mathcal{J_{\alpha}+ I_{\alpha}V_\alpha})=0$.

Equations \eqref{exactcurrents1} and \eqref{exactcurrents2} are exact within the scattering approach,
Now, in the weakly nonlinear transport regime charge and heat currents can be expanded
around the equilibrium state (defined with $V_{\alpha}=0$ and $\theta_\alpha=0$ for all $\alpha$)
up to second order in powers of the driving fields $V_\alpha$ and $\theta_\alpha$:
\begin{align}\label{currents}
I_\alpha &=\sum_{\beta}G_{\alpha\beta}V_\beta
+\sum_{\beta}L_{\alpha\beta}\theta_\beta
+\sum_{\beta\gamma}G_{\alpha\beta\gamma}V_\beta V_\gamma 
\nonumber \\
&+\sum_{\beta\gamma}L_{\alpha\beta\gamma}\theta_\beta \theta_\gamma
+2\sum_{\beta\gamma}M_{\alpha\beta\gamma}V_\beta \theta_\gamma\,,
\\ %\nonumber
\label{eq_heatcurrent}\mathcal{J}_\alpha &=\sum_{\beta} R_{\alpha\beta}V_\beta
+\sum_{\beta}K_{\alpha\beta}\theta_\beta
+\sum_{\beta\gamma}R_{\alpha\beta\gamma}V_\beta V_\gamma 
\nonumber \\
&+\sum_{\beta\gamma}K_{\alpha\beta\gamma}\theta_\beta \theta_\gamma
+2\sum_{\beta\gamma}H_{\alpha\beta\gamma}V_\beta \theta_\gamma\,.
\end{align}
The linear-response electric conductance is given by,
\begin{equation}\label{eq_G}
G_{\alpha\beta}=\frac{2e^2}{h}\int dE\,  A_{\alpha\beta}(E) \,\left[-\partial_E f(E)\right]\,.
\end{equation}
At very low temperature, the Sommerfeld expansion to leading order in $k_B T/E_F$  yields the simple expression
$G\simeq (2e^2/h) \,A_{\alpha\beta}(E_F)$. The linear-response thermoelectric conductance is
\begin{equation}\label{eq_L}
L_{\alpha\beta}=\frac{2e}{hT}\int dE (E-E_F) A_{\alpha\beta}(E)\left[-\partial_E f(E)\right]\,,
\end{equation} 
which reduces to $L_{\alpha\beta}=(2e/hT)(\pi^2 k_B^2/3)\partial_E A_{\alpha\beta}(E)|_{E=E_F}$
after the Sommerfeld expansion is applied. The linear-response
heat current in Eq.~\eqref{eq_heatcurrent} is given by the 
electrothermal conductance
\begin{equation}\label{eq_R}
R_{\alpha\beta}=\frac{2e}{h}\int dE (E-E_F) A_{\alpha\beta}(E)\left[-\partial_E f(E) \right]\,,
\end{equation}
and the thermal conductance
\begin{equation}\label{eq_K}
K_{\alpha\beta}=\frac{2}{h}\int dE \frac{(E-E_F)^2}{T}A_{\alpha\beta}(E)\left[-\partial_E f(E) \right]\,,
\end{equation}
where analytical expressions can be obtained for these conductances considering again the Sommerfeld expansion:
$R_{\alpha\beta}=(2e/h)(\pi^2 k_B^2 T^2/3)\partial_E A_{\alpha\beta}(E)|_{E=E_F}$ and $K_{\alpha\beta}=(2/h)(\pi^2 k_B^2 T/3) A_{\alpha\beta}(E_F)$. Note that the nondiagonal transport coefficients
produce electric current from a temperature difference (the thermoelectric conductance $L_{\alpha\beta}$)
or heat current from a voltage difference (the electrothermal conductance $R_{\alpha\beta}$).
Due to reciprocity, both effects are connected: $R_{\alpha\beta}=TL_{\alpha\beta}$.

The linear response coefficients in Eqs.~\eqref{eq_G}, \eqref{eq_L}, \eqref{eq_R}
and \eqref{eq_K} are evaluated at equilibrium and consequently $G_{\alpha\beta}$, $L_{\alpha\beta}$, $R_{\alpha\beta}$, and $K_{\alpha\beta}$ are independent
of the screening potential $U$.
On the contrary, the leading-order nonlinearities (see below) do depend on the applied electrical and thermal shifts through the electrostatic potential $U$. Indeed, this dependence is responsible for rectification effects in both, electrical and heat currents.
The nonlinear transport coefficients for transport of charge were obtained in Ref. \onlinecite{san12}.
We here find the nonlinear coefficients for the {\em heat} current:
\begin{widetext}
\begin{subequations}
\label{nonlinear}
\begin{align}
& R_{\alpha\beta\gamma}= \frac{e^2}{h} \int dE \partial_E f(E) \Biggr\{\delta_{\alpha\gamma} A_{\alpha\beta} +\delta_{\alpha\beta} A_{\alpha\beta}
- (E-E_F)\left(\frac{\partial A_{\alpha\beta}}{\partial eV_\gamma}+\frac{\partial A_{\alpha\gamma}}{\partial eV_\beta}\right)-
\delta_{\beta\gamma} \left[(E-E_F)\frac{\partial A_{\alpha\beta}}{\partial E}+A_{\alpha\beta} \right]\Biggr\}\,,
\label{Rcoeff} 
\\ 
&K_{\alpha\beta\gamma}=\frac{1}{h} \int dE \partial_E f(E) \frac{(E-E_F)^2}{T}\Biggr\{\left(\frac{\partial A_{\alpha\beta}}{\partial \theta_\gamma} +\frac{\partial A_{\alpha\gamma}}{\partial \theta_\beta}\right)+ \delta_{\beta\gamma} \left[\frac{(E-E_F)}{T}\frac{\partial A_{\alpha\beta}}{\partial E}+ \frac{A_{\alpha\beta}}{T}\right]\Biggr\} \,, \label{Kcoeff}
\\ 
 & H_{\alpha\beta\gamma}=\frac{e}{h}\int dE  \partial_E f(E) (E-E_F) \Biggr\{
\left( \frac{\partial A_{\alpha\gamma}}{\partial\theta_\beta}+ \frac{E-E_F}{T}\frac{\partial A_{\alpha\beta}}{\partial eV_\gamma}-\delta_{\alpha\gamma}\frac{A_{\alpha\beta}}{T} \right)+\delta_{\beta\gamma}\left[\frac{(E-E_F)}{T} \frac{\partial A_{\alpha\beta}}{\partial E} + \frac{A_{\alpha\beta}}{T} \right]\Biggr\}\,. \label{Hcoeff}
\end{align}
\end{subequations}
\end{widetext}
Equations (\ref{Rcoeff}), (\ref{Kcoeff}), (\ref{Hcoeff}),  are formally the main results of this work. Notably, the nonlinear responses depend not only on $A_{\alpha\beta}$ but also on its change with variations of the set of electrical and thermal shifts $\{ V_{\gamma}, \theta_{\gamma}\}$ due to the screening response of the system.

A reasonable model for the interactions as response to the electrical and thermal biases considers only small deviations away from equilibrium. Hence, the internal potential is expanded as
\begin{equation}\label{eq_u}
U=U_\text{eq} + \sum_\alpha u_\alpha V_\alpha +\sum_\alpha z_\alpha \theta_\alpha\,,
\end{equation}
where $u_\alpha=(\partial U/\partial V_\alpha)_\text{eq}$ and $z_\alpha=(\partial U/\partial \theta_\alpha)_\text{eq}$
are the characteristic potentials. These potential susceptibilities are measures of the internal reaction of the conductor
in response to an electrical and thermal shift applied to contact $\alpha$.
For definiteness, we next consider the case of a homogeneous potential
profile independent of the position, although we emphasize that
the extension to inhomogeneous fields in our general
model is straightforward.

We first evaluate
%In general, the nonlinear terms
%$G_{\alpha\beta\gamma}$, $L_{\alpha\beta\gamma}$, $M_{\alpha\beta\gamma}$ and 
%higher order coefficients depend on the characteristic potentials $u$, and $z$.
the total charge $q$ of the conductor, which has two contributions:
\textit{(i)} $q_{\rm bare}$ corresponding to the {\em bare} charge injected from lead $\alpha$ and \textit{(ii)} the {\em screening} charge denoted with $q_{\rm scr}$. The latter corresponds to the charge that builds up inside the 
conductor in response to the injected charges.\cite{but93}
The bare charge $q_{\rm bare}$ injected from lead $\alpha$ originates from a voltage and a temperature imbalance in that terminal. Therefore, $q_\text{bare}$ is decomposed into the particle injectivity $\nu^{p}_{\alpha}(E)$ contribution\cite{but93}  and  the entropic injectivity $\nu^{p}_{\alpha}(E)$ term:\cite{san12}
\begin{eqnarray}\label{eq_dpde}
\nu^{p}_\alpha(E)&=&\frac{1}{2\pi i}\sum_{\beta}
\Tr\left[ s^\dagger_{\beta\alpha}\frac{d s_{\beta\alpha}}{dE} \right]\,,\\ \,
\nu^{e}_\alpha(E)&=&\frac{1}{2\pi i}\sum_{\beta}\Tr\left[\frac{E-E_F}{T}
 s^\dagger_{\beta\alpha}\frac{d s_{\beta\alpha}}{dE} \right]\,.\label{eq_dpde2}
\end{eqnarray}
Importantly, the contribution to $q_{\rm bare}$ due to a temperature shift can be either positive or negative
depending on whether the energy of the carriers is above or below $E_F$. This notorious feature is related with the fact that 
a heat addition or removal depends on whether the carrier energy $E$ is larger or smaller than $E_F$.\cite{hum02}
Then, the factor $(E-E_F)/T$ represents the entropy transfer
associated to the additional carrier into the conductor.
We also note that in Eq.~\eqref{eq_dpde2} we write $E_F$ instead of the
chemical potential $\mu(T)$, which is in general temperature dependent if the reservoir's charge density
is assumed to be fixed.\cite{ashcroft} However, in experimentally relevant situations one externally fixes
the electrochemical potential using, e.g., voltage sources.
Obviously, $\mu\simeq E_F$ in the limit of very low temperatures.
Finally, the total accumulation or depletion bare charge imbalance becomes
\begin{equation}
q_\text{bare}=e\sum_\alpha (D^{p}_\alpha eV_\alpha+D^{e}_\alpha \theta_\alpha )\,,
\end{equation}
where $D_\alpha^{p}=-\int dE \nu^{p}_\alpha (E)\partial_E f$, and $D_\alpha^{e}=-\int dE \nu^{e}_\alpha (E)\partial_E f$ represent the integrated  particle and entropic injectivities around the Fermi energy. 

The screening charge is calculated from the response of the internal potential, $\Delta U=U-U_\text{eq}$,
away from the equilibrium state $U_\text{eq}$. We consider the random phase approximation, in which case
the screening charge is proportional to the Lindhard function $\Pi$, $q_\text{scr}=e^2\Pi \Delta U$, which in the long wavelength limit becomes $\Pi=\int dE D(E) \partial_E f$,\cite{smi83}  with $D=D(E_F)$ the conductor density of states. 
The set of equations  for the characteristic potentials is closed using the Poisson equation. Thus, we relate the out-of-equilibrium net charge $\delta q=q-q_\text{eq}$ to $\Delta U=U-U_\text{eq}$  through $\nabla^2 \Delta U=-4\pi \Delta q$.
By employing Eq.\ \eqref{eq_u} and the fact that $V_\alpha$ and $\theta_\alpha$ shifts are independent,
we find a set of equations for the electrical and thermal characteristic potentials:
\begin{align}
-\nabla^2 u_\alpha + 4\pi e^2 \Pi u_\alpha &= 4\pi e^2 D^{p}_\alpha \,,\\
-\nabla^2 z_\alpha + 4\pi e^2 \Pi z_\alpha &= 4\pi e D^{e}_\alpha\,.
\end{align}

It is computationally useful to aply the WKB approximation to the nonlinear coefficients Eqs.~(\ref{Rcoeff}), (\ref{Kcoeff}) and (\ref{Hcoeff}). Notice that this approach has the same range of validity than the long wavelength limit taken above.
Therefore,  we can make the replacement $\delta/\delta U\to -e\partial/\partial E$ in Eqs.~(\ref{Rcoeff}), (\ref{Kcoeff})
and (\ref{Hcoeff}) and the voltage and temperature derivatives, $\partial_{\theta_\gamma} A_{\alpha\beta}$ and $\partial_{V_\gamma} A_{\alpha\beta}$,
 are calculated once the characteristic potentials are known since
\begin{eqnarray}\label{replacement}
&&\partial_{\theta_\gamma} A_{\alpha\beta}=z_\gamma\delta A_{\alpha\beta}/\delta U\to -e z_\gamma\partial_E A_{\alpha\beta}\,,\\ \nonumber
&&\partial_{V_\gamma} A_{\alpha\beta}=u_\gamma\delta A_{\alpha\beta}/\delta U\to -e u_\gamma\partial_E A_{\alpha\beta}\,.
\end{eqnarray}
Thus, Eqs.~(\ref{Rcoeff}), (\ref{Kcoeff}) and (\ref{Hcoeff})  become
\begin{widetext}
\begin{subequations}\label{coefficients2}
\begin{align}
&R_{\alpha\beta\gamma}= \frac{e^2}{h} \int dE \partial_E f(E) \Biggr\{\delta_{\alpha\gamma} A_{\alpha\beta} +\delta_{\alpha\beta} A_{\alpha\beta}+ (E-E_F)\left(\frac{\partial A_{\alpha\beta}}{\partial E} u_\gamma+\frac{\partial A_{\alpha\gamma}}{\partial E} u_\beta\right)-\delta_{\beta\gamma} \left[(E-E_F)\frac{\partial A_{\alpha\beta}}{\partial E}+A_{\alpha\beta} \right]\Biggr\}\,,
\\
&K_{\alpha\beta\gamma}=-\frac{1}{h} \int dE \partial_E f(E)  \frac{(E-E_F)^2}{T}\Biggr\{\left(e\frac{\partial A_{\alpha\beta}}{\partial E } z_\gamma +e\frac{\partial A_{\alpha\gamma}}{\partial E} z_\beta\right)-\delta_{\beta\gamma} \left[\frac{(E-E_F)}{T}\frac{\partial A_{\alpha\beta}}{\partial E}+ \frac{A_{\alpha\beta}}{T}\right]\Biggr\} \,,
\\
 &H_{\alpha\beta\gamma}=\frac{-e}{h}\int dE \partial_E f(E)  (E-E_F) \Biggr\{\left( e\frac{\partial A_{\alpha\gamma}}{\partial E} z_\beta+ \frac{E-E_F}{T}\frac{\partial A_{\alpha\beta}}{\partial E} u_\gamma+\delta_{\alpha\gamma}\frac{A_{\alpha\beta}}{T} \right)-\delta_{\beta\gamma}\left[\frac{(E-E_F)}{T} \frac{\partial A_{\alpha\beta}}{\partial E} + \frac{A_{\alpha\beta}}{T} \right]\Biggr\} \,.
\end{align}
\end{subequations}
\end{widetext}
In the low temperature limit and for a two terminal device ($\alpha=1,2$), the number of nonlinear conductances in Eq.\ \eqref{coefficients2} can be greatly reduced because the electrothermal and thermal conductances for contact $\alpha=1$ can be expressed in terms of the  characteristic potentials $u_{1,2}$, and $z_{1,2}$ and the total transmission $\mathcal{T}(E)$ as
\begin{subequations}
\begin{align}
&R_{111}\!\approx\frac{e^2}{h}\!\!\left[-\mathcal{T}(E)\!+\!\frac{\pi^2 k_B^2 T^2}{3}\! (1-2u_1)\! \frac{\partial^2 \mathcal{T}(E)}{\partial E^2}\!\right]\!\Biggr |_{E=E_F}\,,
\\
&K_{111}\!\approx\frac{e}{h}\frac{\pi^2 k_B^2 T}{3} \left[ 2 z_1\frac{\partial \mathcal{T}(E)}{\partial E}\!-\!\frac{\mathcal{T}(E)}{eT}\!\right]\!\Biggr|_{E=E_F}\,,
\\
&H_{111}\!\approx\frac{e^2}{h}\frac{\pi^2 k_B^2 T^2}{3} \left[ z_1 \frac{\partial^2 \mathcal{T}(E)}{\partial E^2}\!+\!u_1\frac{1}{eT}\frac{\partial \mathcal{T}(E)}{\partial E}\!\right]\!\Biggr|_{E=E_F}\,,
\end{align}
\end{subequations}

\section{Quantum dot} 
In order to illustrate the general formalism described in the previous section,
we now investigate the nonlinear heat transport for a paradigmatic mesoscopic system: an 
interacting quantum dot. Recent experimental findings support 
interesting nonlinear thermoelectric effects in quantum dots.\cite{sve12} 
We theoretically model the quantum dot system with a single localized level with energy $E_d$.
The dot is attached to two reservoirs via tunneling barriers as shown in Fig. \ref{fig:1}. 
We model such rates with energy independent constants $\Gamma_1$ and $\Gamma_2$,
resulting in a total dot level  broadening given by $\Gamma=\Gamma_1+\Gamma_2$.
 We treat the electrostatic interaction within a mean-field description. 
 Under these considerations, the heat current from contact $1$ is calculated from 
\begin{equation}\label{eq_Jdot}
\mathcal{J}_{1}=\frac{2}{h}\int dE (E-E_F-eV_1) \mathcal{T}(E) \left[f_1(E)-f_2(E)\right]\,,
\end{equation}
where 
\begin{equation}
\mathcal{T}=\frac{4\Gamma_1\Gamma_2}{(E-E_d-eU)^2+\Gamma^2}\,,
\end{equation}
is the corresponding Breit-Wigner transmission line shape for the dot level as a function of the internal potential $U$.
The dot charge reads,
%This potential  has to be self-consistently obtained from the Poisson equation that relates the total dot charge
\begin{equation}\label{q_dot}
q_d=\frac{e}{\pi }\int dE\frac{\Gamma_1 f_1(E)+
\Gamma_2 f_2(E)}{(E-E_d-e U)^2 +\Gamma^2 } \,.
\end{equation}
%to nonequilibrium variation of the dot potential $\Delta U$.
Within a mean-field treatment of interactions,
we solve the discrete version of the Poisson equation by introducing a geometrical capacitance 
 $C$ which connects electrically the dot to an external gate terminal $V_g$ controlling the level position. 
 Then, $U$ is determined from
 \begin{equation}\label{eq_pois}
 \delta q_d=q_d-q^{\rm eq}_d= C(U-V_g)\,,
 \end{equation}
 where $\delta q_d$ represents the charge excess due to electrical and thermal applied biases
 while $q^{\rm eq}_d$ is the equilibrium charge calculated from Eq. (\ref{q_dot}) by setting $f_1=f_2=f$.

Once the internal dot potential $U$ is obtained, Eq.\eqref{eq_Jdot} can be integrated and yields
 \begin{eqnarray}\label{heatexact}%\nonumber
 \mathcal{J}_{1}\!=\!\frac{4\Gamma_1\Gamma_2}{\Gamma\hbar}{\rm Im}\Biggr\{\!\!\beta_1\xi_1\!
\Biggr[\! \Psi\left(\frac{1}{2}\!+\!i\frac{\xi_2}{\pi}\right)\!\!-\!\!\Psi\left(\frac{1}{2}+\frac{i\xi_1}{\pi}\!\right)\!\Biggr]\!\Biggr\}
\end{eqnarray}
where $\Psi$ is the digamma function with arguments $\xi_{1(2)}=\beta_{1(2)}[\tilde{E}_d-i\Gamma-\mu_{1(2)}]$ where $\beta_{1(2)}=1/k_B T_{1(2)}$. Importantly, the dot level is renormalized by the interactions: $\tilde{E}_d=E_d+eU$.

Equation \ref{heatexact} is the exact expression for the heat current 
when Coulomb interaction is considered within a mean-field description. It is to be compared
to the voltage-temperature expansion obtained from 
the nonlinear heat transport formalism discussed in Sec.\ II.
To do so, we calculate the heat transport conductances in terms of the characteristic potentials
and expand Eq. (\ref{q_dot}) to leading order in $V_\alpha$, $\theta_\alpha$ and $U$. We find
\begin{equation}\label{eq_dotcharge}
\delta q_d =  e^2 D^{p}_1 V_1+ e^2 D^{p}_2 V_2 + e 
D^{e}_{1}  \theta_1 + e D^{e}_{2} \theta_2 - e^2 D U\,,
\end{equation}
where
\begin{eqnarray}
D^{p}_{\alpha}&=&\frac{-\Gamma_\alpha}{\pi}\int dE \frac{\partial_E f(E)}{(E-E_d)^2 +\Gamma^2}  \,, \label{eqdpde}
\\
D^{e}_{\alpha}&=&\frac{-\Gamma_\alpha}{\pi}\int dE  \frac{E-E_F}{T}\frac{\partial_E f(E)}{(E-E_d)^2 +\Gamma^2}\,.\label{eqdpde2}
\end{eqnarray}
Notice that Eqs.\ \eqref{eqdpde} and \eqref{eqdpde2} are the integrated particle and entropic injectivities
within a Breit-Wigner representation of the dot scattering matrix.
Using Eq.\ \eqref{eq_pois} and \eqref{eq_dotcharge} we find the dot internal potential
\begin{equation}\label{internal_U}
U=\frac{e^2 D^{p}_1 V_1+ e^2 D^{p}_2 V_2 + 
e D^{e}_{1}  \theta_1 + e D^{e}_{2} \theta_2 +C V_g}{C+e^2 D}\,,
\end{equation}
from which the characteristic potentials follow,
\begin{equation}\label{equz}
u_{1(2)}=\frac{e^2 D^p_{1(2)}}{C+e^2 D},\,\,u_g=\frac{C}{C+e^2 D},\,\,
z_{1(2)}=\frac{e D^e_{1(2)}}{C+e^2 D}\,.
\end{equation}
Note that while $u_1 +u_2 +u_g = 1$ because of gauge invariance \cite{but93} such a sum rule is not satisfied
by the $z$ potentials .

\section{Heat rectification effects}
Rectification effects occur in systems where the functional
dependence of current versus the driving field
(voltage or temperature shift) departs from being linear
due to quadratic (nonlinear) transport responses.
We now investigate nonlinearities in the heat flux--voltage
and heat--temperatures characteristics of a quantum
dot. We consider the charge neutral limit ($C=0$),
in which case strong interactions renormalize the energy
level to maintain a fixed charge inside the dot.
This limit is relevant in many experimental situations.
 \begin{figure}
  \centering
 \includegraphics[width=0.45\textwidth, angle=-90, clip]{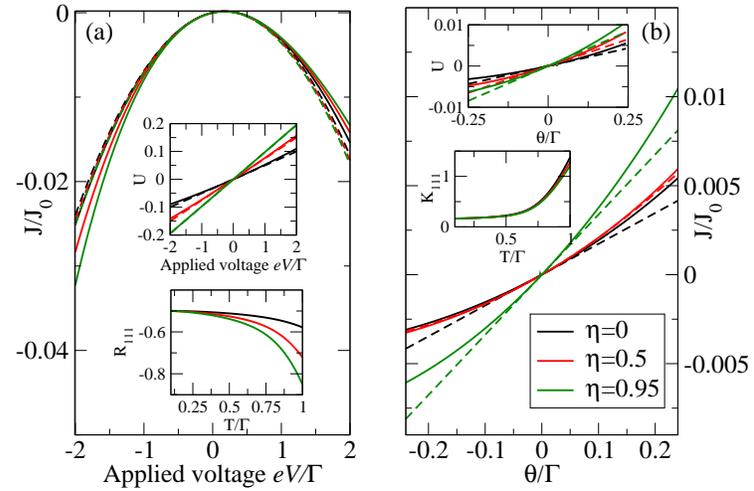}
  \caption{(Color online)  Heat current $\mathcal{J}$ (in units of $\mathcal{J}_0=2\Gamma_1\Gamma_2/h$)
for a quantum dot system with $\mu_1=E_F+eV$, 
$\mu_2=E_F$, $\theta_1=T+\theta$, and 
$\theta_2=T$ for three different values
of the tunneling asymmetry $\eta$ and $E_d=10\Gamma$ with $T=0.5\Gamma$.
(a) $\mathcal{J}$--$V$ characteristics (full lines) for $\theta=0$ along with 
the leading-order nonlinearity $\mathcal{J}\simeq R_{11}V+R_{111} V^2$ (dashed lines)
for  the $\eta$ values indicated in the right panel. 
Upper inset: we show with solid lines
the self-consistent potential $U$ as a function of $eV/\Gamma$.
Dotted lines correspond to the approximated values $U\approx u V=\Gamma_1/\Gamma$
to leading order in an expansion in powers of $V$.
Lower inset: temperature dependence of $R_{111}$.
(b) $\mathcal{J}$--$\theta$ characteristics (full lines) for $V=0$ along with
the leading-order nonlinearity $\mathcal{J}\simeq K_{11} \theta+K_{111}\theta^2$ (dashed lines)
for the indicated values of $\eta$.
Upper inset: we show with solid lines the self-consistent potential $U$ as a function of
the thermal shift. Lower inset: temperature dependence of
the nonlinear thermal conductance $K_{111}$.}
  \label{fig:2}
\end{figure}

In the isothermal case ($T_1=T_2$), we take $V_1=V$ and $V_2=0$.
Then, to leading order in voltage the characteristic potential becomes
$u=\partial U/\partial V=(1+\eta)/2$ with 
$\eta=(\Gamma_1-\Gamma_2)/\Gamma$ the tunneling asymmetry.\cite{san04}
In the upper inset of Fig.\ \ref{fig:2}(a) we show the exact dot potential $U$
obtained from a numerical, self-consistent calculation
of Eq.\ \eqref{q_dot} compared to its leading-order value $U=uV$.
We observe that the approximation
is rather good even for voltages larger than the level broadening $\Gamma$.
The heat current $\mathcal{J}\equiv \mathcal{J}_1 $ is shown in the main part of Fig.\ \ref{fig:2}
as a function of voltage. At low voltages, the relation is linear but is not
clearly visible since the quadratic-in-$V$ term largely dominates (dashed lines).
We recall that at low temperatures $\mathcal{J}\simeq R_{111}V^2$  represents a Joule heating term,
which is the main source of heat release in a voltage-biased terminal.
In the lower inset of Fig.\ \ref{fig:2}(a) we plot
the nonlinear electrothermal conductance
as a function of the background temperature.
Interestingly, $R_{111}$ increases in magnitude
more quickly as the tunneling asymmetry
increases. 

In the isoelectric case ($\mu_1=\mu_2$),
we take $\theta_1=\theta$ and $\theta_2=0$.
The characteristic potential determined from the entropic
injectivities is $z=\partial U/\partial \theta=D_1^e/[2e (D_1^p+D_2^p)]$.
Thus, the dot  potential reduces to $U=z\theta$
to leading order in $\theta$. We compare in the 
upper inset of Fig.\ \ref{fig:2}(b) the validity of this approximation
with the exact calculation of $U$. We find that for thermal shifts
much smaller than $\Gamma$, the relation between $U$
and $\theta$ stays linear. In the main part of Fig.\ \ref{fig:2}(b) 
we depict the heat current as function of $\theta$
for various values of the tunneling asymmetry.
Since we select small values of $\theta$,
the agreement between the weakly nonlinear
expressions and the full calculation is excellent.
In the lower inset of Fig.\ \ref{fig:2}(b)
we show the nonlinear thermal conductance
$K_{111}$ as a function of the background temperature $T$.
Similarly to the isothermal case, the rectifying thermal conductance
increases with higher values of $T$.

\section{Nonlinear Peltier effect}
The Peltier effect accounts for the generated heat
in a given terminal caused by the flow of a unit charge
through the conductor, keeping all terminals at the same temperature.
More precisely, the two-terminal Peltier coefficient is defined at linear response
as the ratio between heat and electrical currents,
\begin{equation} \label{eq_pi0}
\Pi_0=\frac{\mathcal{J}_1}{\mathcal{I}_1}=\frac{R_{11}}{G_{11}}\,,
\end{equation}
when $\theta_\gamma=0$ for all $\gamma$.
Thus, the Peltier coefficient can be viewed as the analogue to the thermopower $S_0$
(a voltage generation in response to a thermal gradient under the condition
of zero net current). In fact, both coefficients are connected via the Kelvin relation $\Pi_0= T S_0$.
The Peltier effect informs us about how efficiently electrical currents are converted into heat currents whereas the Seebeck effect quantifies the transformation of waste heat into useful electricity.

Our goal here is to generalize Eq.\ \eqref{eq_pi0} to the nonlinear case.
We consider the following expression,
\begin{equation}\label{eq_peltiernl}
\Pi_\alpha=\frac{\mathcal{J}_\alpha}{\mathcal{I}_\alpha}\Biggr |_{\{\theta_\gamma=0\}}\,.
\end{equation}
To calculate the Peltier coefficient in the weakly nonlinear regime we employ the second order expansion in terms of electrical and temperature gradients for both the electrical and heat currents. For a two-terminal device an isothermal electrical current $\mathcal{I}_1=\mathcal{I}$ is driven by a bias voltage drop $V$,
\begin{equation}\label{electricalPeltier}
\mathcal{I}= G_{11} V+ G_{111} V^2+\mathcal{O}(V^3)\,.
\end{equation} 
Inverting this expression, we substitute the voltage $V =\mathcal{I}/G_{11}-\mathcal{I}^2 G_{111}/G_{11}+\mathcal{O}(\mathcal{I}^3)+\cdots$ into the heat current,
\begin{equation}\label{heatPeltier}
\mathcal{J}_{1}= R_{11}V+ R_{111}V^2+\mathcal{O}(V^3)\,.
\end{equation}
 yielding
\begin{equation}\label{eq_Jp}
\mathcal{J}_1=\frac{R_{11}}{G_{11}}  \mathcal{I}+ \frac{1}{G_{11}}\left(\frac{R_{111}}{R_{11}}-\frac{G_{111}}{G_{11}}\right) \mathcal{I}^2 + \mathcal{O}(\mathcal{I}^3)\,,
\end{equation}
We insert Eq.\ \eqref{eq_Jp} in Eq.\ \eqref{eq_peltiernl} and find 
\begin{equation}\label{eq_dimPi}
\Pi=\Pi_0\left[1+\frac{1}{G_{11}}\left(\frac{R_{111}}{R_{11}}-\frac{G_{111}}{G_{11}}\right)\mathcal{I}+\cdots\right]
\end{equation}
where we have set $\Pi\equiv \Pi_1$. Now, the second term in the r.h.s of Eq. (\ref{eq_dimPi})  measures the deviations of $\Pi$ from the linear-response value, $\Pi_0$. Interestingly enough, nonlinear contributions are dictated by a ratio between the nonlinear to linear thermal and electrical conductances. In other words, the conversion efficiency for electric currents into heat flow
is given by the relative strength of the nonlinear conductances to the linear ones
and by the difference between the heat and the electrical properties of the conductor.

\begin{figure}
  \centering
 \includegraphics[width=0.4\textwidth, angle=-90,clip]{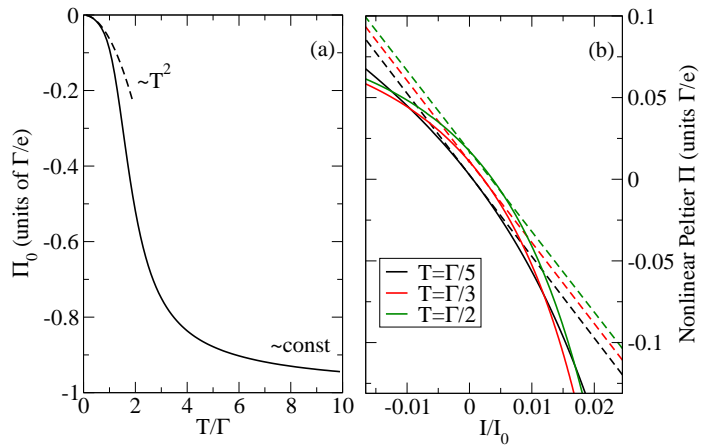}
  \caption{(Color online) (a) Linear-response Peltier coefficient $\Pi_0$ for a current biased quantum dot
  at $E_d=10\Gamma$. We show both the low and high temperature limits. (b) Full Peltier coefficient $\Pi$
  including contributions beyond linear response
 for three different background temperatures for $\eta=0.5$. We show with colored dotted lines the leading-order
 expansion calculated from Eq.\ \eqref{eq_dimPi}.}
  \label{fig:3}
\end{figure}

We show in Fig.\ \ref{fig:3}(a) the linear Peltier coefficient for a two-terminal
quantum dot as described in Sec.\ II taking $G_{111}=(e^3/h) (1-2 u_1) \partial_E\mathcal{T}(E)|_{E=E_F}$.\cite{but93,chr96}  As the background $T$ increases, $\Pi_0$ quickly
enhances following a $T^2$ law and then saturates for temperatures much larger
than the resonance broadening $\Gamma$.
The first fact can be understood from a Sommerfeld expansion of $R_{11}$ and $G_{11}$.
The former goes as $T^2$ whereas the latter is independent of temperature to leading
order in $T$. Finally, the saturation value can be deduced by replacing the Lorentzian
shape of the Breit-Wigner resonance with a delta-like peak, an approximation valid
in the high temperature regime ($T\gg\Gamma$). Then, one finds $\Pi_0=-(E_0-E-F)/e$.
Note that this expression obeys the Kelvin relation $\Pi_0=T S_0$ since
the linear-response thermopower for the same system
is $S_0=-(E_0-E-F)/{eT}$ at high temperatures.\cite{san12}
\begin{figure}[t]
  \centering
 \includegraphics[width=0.4\textwidth, clip]{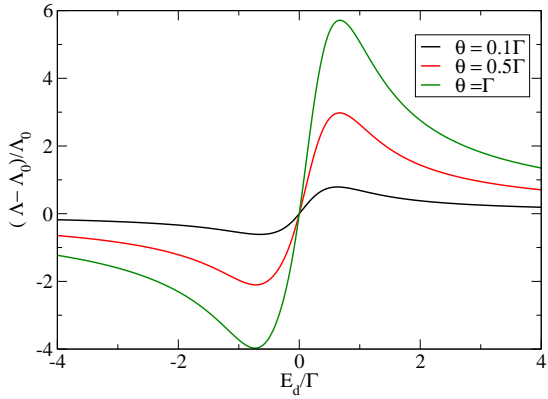}
  \caption{(Color online) Nonlinear departures of the Wiedemann-Franz law
  as expressed by Eq.\ \eqref{eq_ratio}. Nonlinear transport responses are calculated
  for a two-terminal quantum dot with voltage bias $V=0.25\Gamma$,
  background temperature $T=0.1\Gamma$ and different
  values of thermal shift $\theta$ as a function of the dot level position $E_d$.}
  \label{fig:4}
\end{figure}

In the nonlinear regime, $\Pi$ depends on the nonlinear responses according to
Eq.\ \eqref{eq_dimPi} and a self-consistent calculation is thus needed.
We depict in Fig.\ \ref{fig:3}(b) the exact value of $\Pi$ obtained from
a numerical calculation of Eqs.\ \eqref{exactcurrents1}, and \eqref{exactcurrents2} in the mean-field approach for interactions
(full lines), together with the weakly nonlinear result expressed by Eq.\ \eqref{eq_dimPi}.
To leading order in $\mathcal{I}/\mathcal{I}_0$, where $\mathcal{I}_0=(2e^2/h)\Gamma_1\Gamma_2/\Gamma$,
the Peltier coefficient shows a linear dependence of the current, in agreement
with Eq.\ \eqref{eq_dimPi}. Furthermore, $\Pi$ increases for increasing $\mathcal{I}$
but only in the low current regime. At higher currents, highly nonlinear terms
start to contribute and deviations from the linear regime are observed.

\section{Wiedemann-Franz law} 
The Wiedemann-Franz law for metals establishes that the ratio between the linear-response
thermal ($K$) and electrical ($G$) conductances 
is proportional to the temperature ($T$) with a proportionality constant given by the Lorenz number,
$\Lambda_0=(\pi^2/3)(k_B/e)^2$.
This relation holds as long as the Sommerfeld expansion is valid.
More concisely, the Wiedemann-Franz law for a mesoscopic system states that
\begin{equation}\label{eq_wl0}
\frac{K}{TG}=\Lambda_0
\end{equation}
which is obviously satisfied in the linear regime since
$K\simeq (2/3h) \pi^2 k_B^2 T \mathcal{T}(E_F)$ and $G\simeq (2e^2/h) \mathcal{T}(E_F)$.

To generalize this result to the nonlinear regime and to test whether the Wiedemann-Franz law is still valid or departs from the Lorenz value we consider the Wiedemann-Franz ratio
\begin{eqnarray}
\Lambda_\alpha=\frac{(\mathcal{J}_{\alpha}/\theta)\Big|_{\{V=0\}}}{T(\mathcal{I}_\alpha/V) \Big|_{\{\theta=0\}}}\,,
\end{eqnarray}
For illustrative purposes we have considered again the case of a conductor attached to two terminals, where only one of the contacts is heated or cooled with thermal difference $\theta$ and electrically biased with voltage $V$; e.g., $\theta_1=\theta$, $\theta_2=0$,  $V_1=V$, and $V_2=0$. Then,
\begin{eqnarray}
\Lambda =\frac{1}{T} \frac{K_{11}+K_{111}\theta+\cdots}{G_{11}+G_{111}V+\cdots}\,.
\end{eqnarray}
Keeping only terms linear in $\theta$ and $V$, the nonlinear analogue to the Wiedemann-Franz ratio is
\begin{eqnarray}\label{eq_lambdanl}
\Lambda =\frac{1}{T}\left( \frac{K_{11}}{G_{11}} - \frac{ K_{11}G_{111} V}{G_{11}^2} +\frac{K_{111}\theta}{G_{11}}+\cdots\right)
\end{eqnarray}\,.
Clearly, Eq.\ \eqref{eq_wl0} reduces to Eq.\ \eqref{eq_lambdanl} if nonlinear terms are neglected.
Departures from the Wiedemann-Franz rule occurring in the nonlinear regime of transport
can thus be quantified with the following magnitude:
\begin{equation}\label{eq_ratio}
\frac{\Lambda-\Lambda_0}{\Lambda_0} =
\frac{K_{111}}{K_{11}}V-\frac{G_{111}}{G_{11}}\theta\,.
\end{equation}

Interestingly, deviations from the Lorenz number are given by the difference between the ratios of nonlinear to linear thermal and electrical conductances.

We calculate Eq.\ \eqref{eq_ratio} for the quantum dot model described
in Sec.\ III for $V=0.25\Gamma$ and $T=0.01\Gamma$ and different
values of $\theta$. We show our results in Fig.\ \ref{fig:4} as a function
of the dot level position $E_d$. We find that departures are stronger for increasing
temperature differences since for increasing $\theta$
the heat transport becomes more nonlinear. Moreover,
when $E_d$ lies around $\Gamma$
above or below the Fermi energy departures are more apparent
because the transmission function changes rapidly for energies
around those points.

\section{Conclusions}
We have investigated the heat flux in a mesoscopic
system beyond linear response. Our scattering approach
includes interactions to take into account the system's
screening response to voltage and temperature shifts
in the reservoirs. We have found exact expressions for
the leading-order nonlinearities in a voltage-temperature
expansion of the heat current and have applied our general
theory to the study of the heat rectification properties
of an interacting quantum dot attached to two terminals.
We have found that the Peltier coefficient acquires higher-order
contributions at moderate electric currents and that
these contributions are stronger for increasing
background temperature. Furthermore, we have investigated
departures from the Wiedemann-Franz law in the weakly nonlinear
regime of transport and have found that the strongest deviations
occur when charge fluctuations dominate, i.e., when
the distance between the dot level and the leads' Fermi energy
is of the order of the resonance broadening. We believe
that our results are of fundamental importance for the
understanding of basic nonlinear heating effects
in quantum conductors.

\section*{Acknowledgments}
Work supported by MINECO Grant No. FIS2011-23526,
the Conselleria d'Educació, Cultura i Universitats
(CAIB) and FEDER. During the completion stage
of this manuscript, we became aware of a closely related work
by J. Meair, and Ph. Jacquod in Ref.\ \onlinecite{mea12}. The difference is that
they are concerned with thermodynamic efficiencies while
we focus on the nonlinear Peltier effect and departures
from the Wiedemann-Franz law.

\end{document}